\documentclass[useAMS,usenatbib]{mn2e}
\usepackage{txfonts}
\usepackage{graphicx}
\usepackage{amssymb}
\usepackage[below]{placeins}
\usepackage{txfonts}
\usepackage{natbib}
\bibpunct{(}{)}{;}{a}{}{,}
\def \arcmin{$^{\prime}$}

\def \xmm{{\emph{XMM-Newton}}}
\def \rosat{{\emph{ROSAT}}}
\def \chandra{{\emph{Chandra}}}

\newcommand{\ion}[2]{#1\,{\sc{#2}}}

\bibpunct{(}{)}{;}{a}{}{,}

\title[Constraints on turbulent pressure in the X-ray halos of giant elliptical galaxies]
{Constraints on turbulent pressure in the X-ray halos of giant elliptical galaxies from resonant scattering}

\author[Werner et al.]{N. Werner$^{1}$\thanks{Chandra/Einstein fellow, E-mail: norbertw@stanford.edu}, 
I. Zhuravleva$^{2}$, E. Churazov$^{2,3}$, A. Simionescu$^{4}$, S. W. Allen$^{1}$,
 \newauthor  W. Forman$^{5}$, C. Jones$^{5}$, J.S. Kaastra$^{6,7}$\\
$^{1}$Kavli Institute for Particle Astrophysics and Cosmology, Stanford University, 452 Lomita Mall/mc~4085, Stanford, CA 94305, USA\\
$^{2}$Max-Planck-Institut f\"ur Astrophysik, Karl-Schwarzschild-Strasse 1, 85741 Garching, Germany \\
$^{3}$Space Research Institute (IKI), Profsoyznaya 84/32, Moscow 117810, Russia\\
$^{4}$Max-Planck-Institut f\"ur Extraterrestrische Physik, Giessenbachstr, 85748 Garching, Germany\\
$^{5}$Harvard-Smithsonian Center for Astrophysics, 60 Garden St., Cambridge, MA 02138, USA\\
$^{6}$SRON Netherlands Institute for Space Research, Sorbonnelaan 2, 3584 CA Utrecht, the Netherlands\\
$^{7}$Sterrenkundig Instituut, Universiteit Utrecht, P.O. Box 80000, 3508 TA Utrecht, the Netherlands\\
}
\begin{document}
\maketitle
\begin{abstract}
The dense cores of X-ray emitting gaseous halos of large elliptical galaxies with temperatures $kT\lesssim0.8$~keV show two prominent \ion{Fe}{xvii} emission features, which provide a sensitive diagnostic tool to measure the effects of resonant scattering. We present here high-resolution spectra of five bright nearby elliptical galaxies, obtained with the Reflection Grating Spectrometers (RGS) on the \xmm\ satellite. The spectra for the cores of four of the galaxies show the \ion{Fe}{xvii} line at 15.01~\AA\ being suppressed by resonant scattering. The data for NGC~4636 in particular allow the effects of resonant scattering to be studied in detail and to prove that the 15.01~\AA\ line is suppressed only in the dense core and not in the surrounding regions. Using deprojected density and temperature profiles for this galaxy obtained with the {\it{Chandra}} satellite, we model the radial intensity profiles of the strongest resonance lines, accounting for the effects of resonant scattering, for different values of the characteristic turbulent velocity. 
Comparing the model to the data, we find that the isotropic turbulent velocities on spatial scales smaller than $\approx$1~kpc are less than 100~km~s$^{-1}$ and the turbulent pressure support in the galaxy core is smaller than 5\% of the thermal pressure at the 90\% confidence level, and less than 20\% at 95\% confidence. 
Neglecting the effects of resonant scattering in spectral fitting of the inner 2~kpc core of NGC~4636 will lead to underestimates of the chemical abundances of Fe and O by $\sim$10--20\%. 
\end{abstract}


\section{Introduction}


The hot intra-cluster medium and the hot halos around giant elliptical galaxies are usually assumed to be optically thin. Although this assumption is valid for most of the emitted X-ray photons, at the energies of the strongest resonant transitions, the hot plasma can be optically thick \citep{gilfanov1987}. The transition probabilities of strong resonance lines are large and, if the column density of the ion along a line of sight is sufficiently high, photons with the energies of these resonance lines will get absorbed and, within a short time interval, reemited in a different direction. Because of the short time between absorption and emission, this process can be regarded as scattering.
Since resonant scattering in clusters and elliptical galaxies will cause the radial intensity profile of an emission line to become weaker in the centre and stronger outside, it can also lead to an underestimate of metal abundances in the dense cores and a corresponding overestimate in the surrounding region. This effect is, however, not expected to be large. \citet{sanders2006} found that metallicities in cluster cores could be underestimated by at most 10\% due to resonant scattering. 

\citet{gilfanov1987} pointed out that since the optical depth $\tau$ in the core of a resonance line depends on the characteristic velocity of small-scale motion, measurements of $\tau$ give important information about the turbulent velocities in the hot plasma. Measuring the level of resonant scattering in clusters and giant elliptical galaxies is thus a good way to constrain the energy in turbulence and the turbulent pressure support. The first constraints on turbulent velocities using resonant scattering were obtained for the Perseus cluster by \citet{churazov2004}. Using XMM-Newton EPIC data, they compared the relative fluxes of the 1s--2p and 1s--3p He-like Fe lines in the core and in an annulus surrounding the core of the Perseus cluster. The expected optical depth of the 6.7~keV 1s--2p \ion{Fe}{xxv} resonance line is much larger than that of the 1s--3p line, therefore the ratio provides information about the level of resonant scattering. \citet{churazov2004} found no evidence for resonant scattering in Perseus, indicating that differential gas motions on scales smaller than $\sim$100~kpc in the core of the cluster must have a range of velocities of at least half of the sound speed. Independently, using the same data, \citet{gastaldello2004} reached similar conclusions. 

\begin{figure}
\includegraphics[width=0.40\textwidth,clip=t,angle=270.]{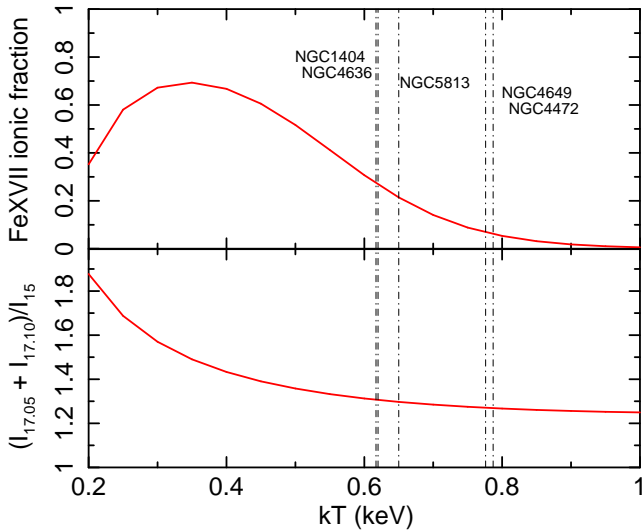}
\caption{Fraction of Fe in the form of \ion{Fe}{xvii} and the theoretical line ratio $(I_{\lambda17.05}+I_{\lambda17.10})/I_{\lambda15.01}$ for an optically thin plasma as a function of the plasma temperature. The vertical lines mark the RGS measured temperatures of the five galaxies in our sample.} 
\label{fractions}
\end{figure}

The most sensitive spectral lines to determine the level of resonant scattering are the \ion{Fe}{xvii} lines at 15.01~\AA\ (2p--3d) and the unresolved blend of the same ion at 17.05 and 17.10~\AA\ (2p--3s). The optical depth is directly proportional to the column density of the ion and the oscillator strength of the given transition. While the oscillator strength of the 15.01~\AA\ line is $f=2.73$, and thus the line is expected to have a relatively large optical depth, the oscillator strength of the 17.05~\AA\ line is $f=0.12$ and the optical depth of this line blend is negligible (the oscillator strength of the 17.10~\AA\ line is of the order of $10^{-8}$)\footnote{http://cxc.harvard.edu/atomdb/WebGUIDE/index.html}. Because of this dramatic difference in expected optical depths, and the fact that both lines originate from the same ion of the same element, the \ion{Fe}{xvii} lines provide, in principle, an excellent diagnostic tool to measure the magnitude of resonant scattering. At the relatively low temperatures of hot halos around giant elliptical galaxies, $\sim$0.6--0.8~keV, both lines are very strong and, in this temperature range, their expected intensity ratios have only a very weak dependence on temperature \citep[see Fig.~\ref{fractions},][]{doron2002}. 
Therefore, any observed spatial dependence of the intensity ratios is unlikely to be due to gradients in temperature or Fe abundance in the hot plasma, making these lines especially clean diagnostic tools.  The \ion{Fe}{xvii} line ratios in extended objects can be measured with the \xmm\ Reflection Grating Spectrometers \citep[RGS,][]{herder2001}. 
\citet{xu2002} reported the first and so far only radial profile of the $(I_{\lambda17.05}+I_{\lambda17.10})/I_{\lambda15.01}$ ratio for the X-ray luminous elliptical galaxy NGC~4636, which indicates the presence of resonant scattering. Attempts to measure resonant scattering in NGC~5044 and M~87, which are hotter and more massive systems with weaker \ion{Fe}{xvii} lines, were not successful \citep{tamura2003,werner2006b}.

Here we follow up on the work by \citet{xu2002}, by reporting the \ion{Fe}{xvii} $(I_{\lambda17.05}+I_{\lambda17.10})/I_{\lambda15.01}$ line ratios measured in the cores of five nearby bright elliptical galaxies observed with \xmm\ RGS: NGC~4636, NGC~5813, NGC~1404, NGC~4649, and NGC~4472. In Sect.~2 we describe the sample and observations; in Sect.~3 we discuss the details of the data analysis; and in Sect.~4 we present the results of the observations. We focus our attention on NGC~4636, for which data are of very high quality. Because NGC~4636 is X-ray bright, has a relatively long exposure and a favorable temperature, for this system we can measure and compare the observed line ratios in the core and surrounding regions. In Sect.~5 we present a model for the radial profile of the 15.01~\AA\ line in NGC~4636 for different values of turbulent velocities, derived using deprojected density and temperature profiles from {\it{Chandra}} data. Comparing this model to the data, we place constraints on the characteristic velocity of isotropic turbulence in the core of the galaxy. Implications of the results are discussed in Sect.~6. 

Throughout the paper, abundances are given with respect to the ``proto-solar values'' by \citet{lodders2003}. 
All errors are quoted at the 68\% confidence level for one interesting parameter ($\Delta C=1$; for C statistics).

\begin{table*}
\begin{center}
\caption{List of the galaxies in the sample, with their distance (luminosity distances, obtained from the NASA/IPAC Extragalactic 
Database, http://nedwww.ipac.caltech.edu/), corresponding linear scale per arcminute, Galactic $N_{\mathrm{H}}$ value \citep{kalberla2005}, average temperature measured by \rosat. The last three columns give the \xmm\ revolution of the observation, and the total and filtered RGS exposure times, respectively.}
\begin{tabular}{lccccccc}
\hline\hline
galaxy     		 &  distance 	& scale      	&$N_{\mathrm{H}}$        	& k$T$	 		& rev.  	& Exp. time 	& Clean time 	\\
		 	& Mpc       		&  kpc/arcmin	& $10^{20}$~cm$^{-2}$ 	& keV 	          	&        	&      ks		&    ks		\\
\hline	
NGC 4636    	&	17.5		&     5.1    		&1.90				&  0.55$^\ast$    	&    197     &    64406   	&  58205 		\\
NGC 5813    	&	29.9		&     8.7    		& 4.37				&  0.52$^\dagger$    	&    1029	&   36708 	 	&  29899 		\\
NGC 1404    	&	25.5		&     7.4    		& 1.51				&  0.60$^\ast$   	&     1033	&    55032  	&  17231		\\
NGC 4649    	&      19.8		&     5.8    		& 2.1					&  0.78$^\ast$    	&     196	&    54210 	& 46096 		\\
NGC 4472    	&      18.4		&     5.4    		& 1.53				&  0.88$^\ast$    	&     744	&    111033 	& 81850		\\
\hline
\noindent $^\ast$ \citet{osullivan2003} \\
\noindent $^\dagger$ \citet{reiprich2002}

\label{obslog}
\end{tabular}
\end{center}
\end{table*}

\section{Sample and XMM-Newton observations}

\begin{figure}
\includegraphics[width=0.47\textwidth,clip=t,angle=0.]{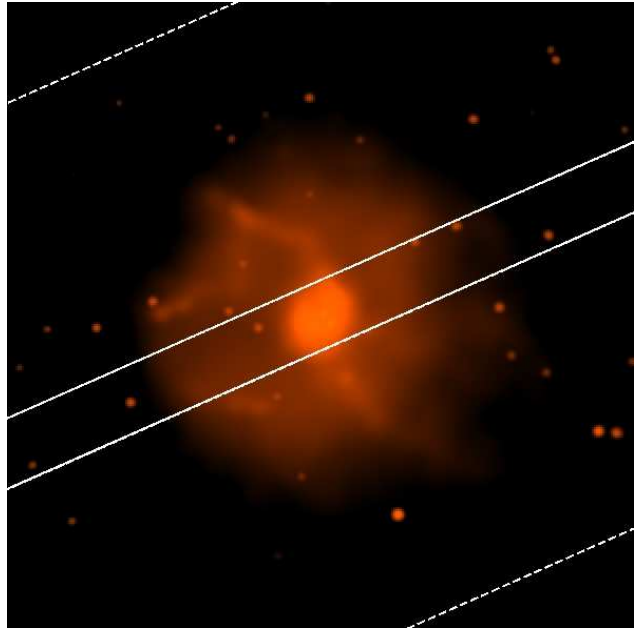}
\caption{{\it{Chandra}} image of NGC~4636 with the RGS extraction regions over-plotted. The central extraction region is 0.5\arcmin\ wide (indicated by full lines), the two extraction regions next to the core (between the full and dashed lines) are 2.25\arcmin\ wide. The image was extracted in the 0.5--2.0~keV band, cleaned and adaptively smoothed. } 
\label{fig:ngc4636}
\end{figure}

\begin{figure*}
\begin{minipage}{0.48\textwidth}
\includegraphics[width=0.85\textwidth,clip=t,angle=0.]{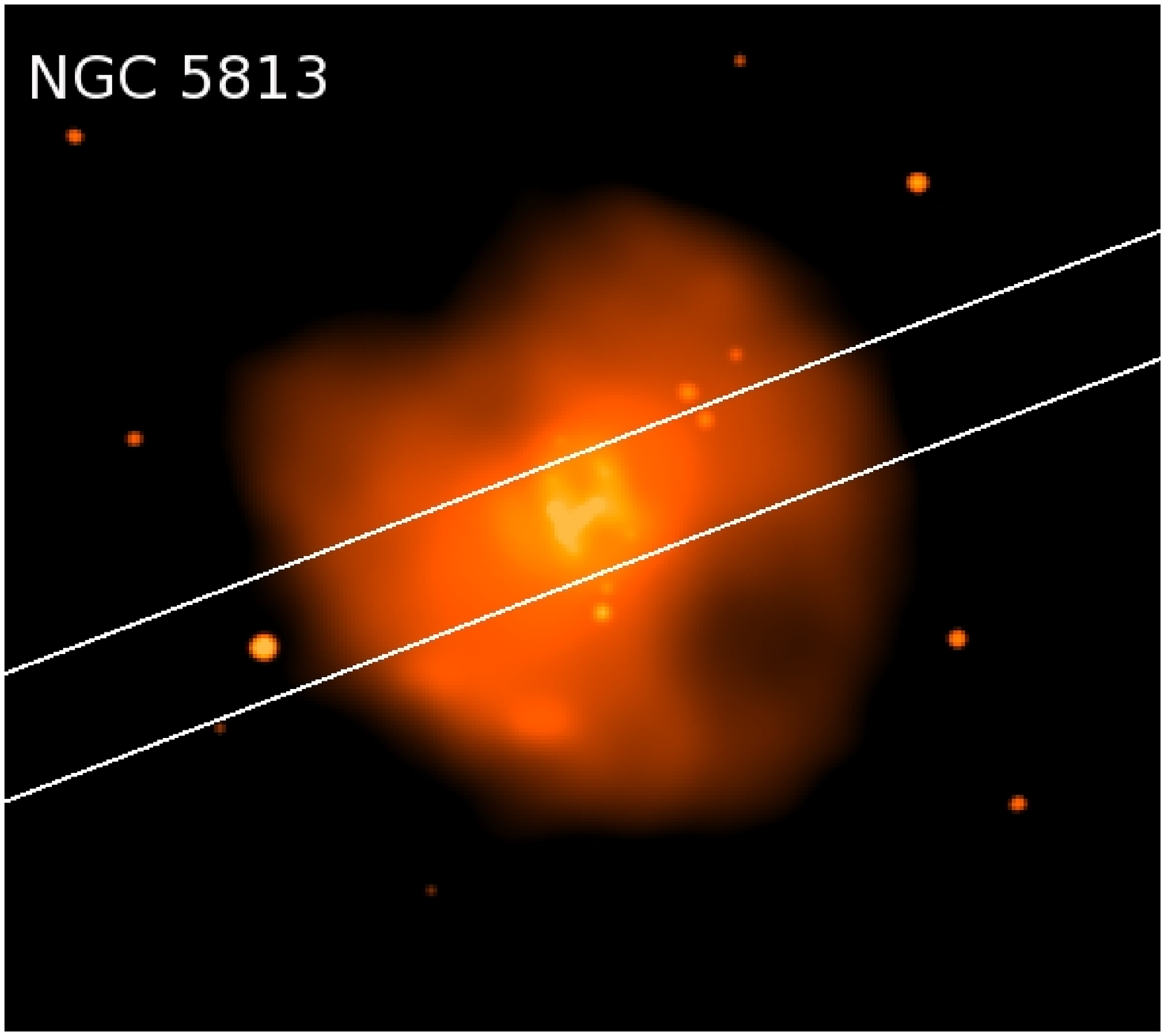}
\end{minipage}
\begin{minipage}{0.48\textwidth}
\includegraphics[width=0.85\textwidth,clip=t,angle=0.]{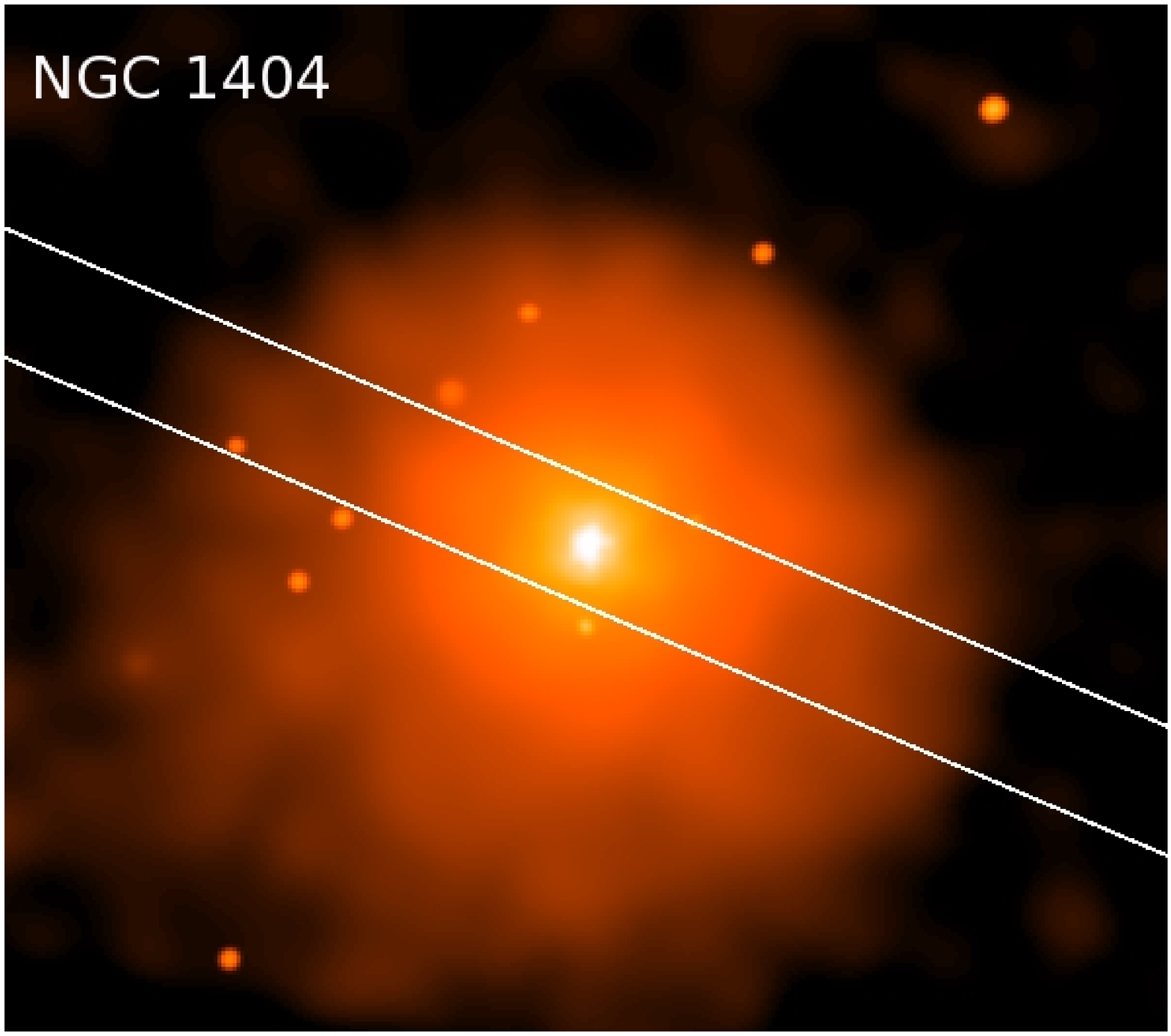}
\end{minipage}
\vspace{2mm}

\begin{minipage}{0.48\textwidth}
\includegraphics[width=0.85\textwidth,clip=t,angle=0.]{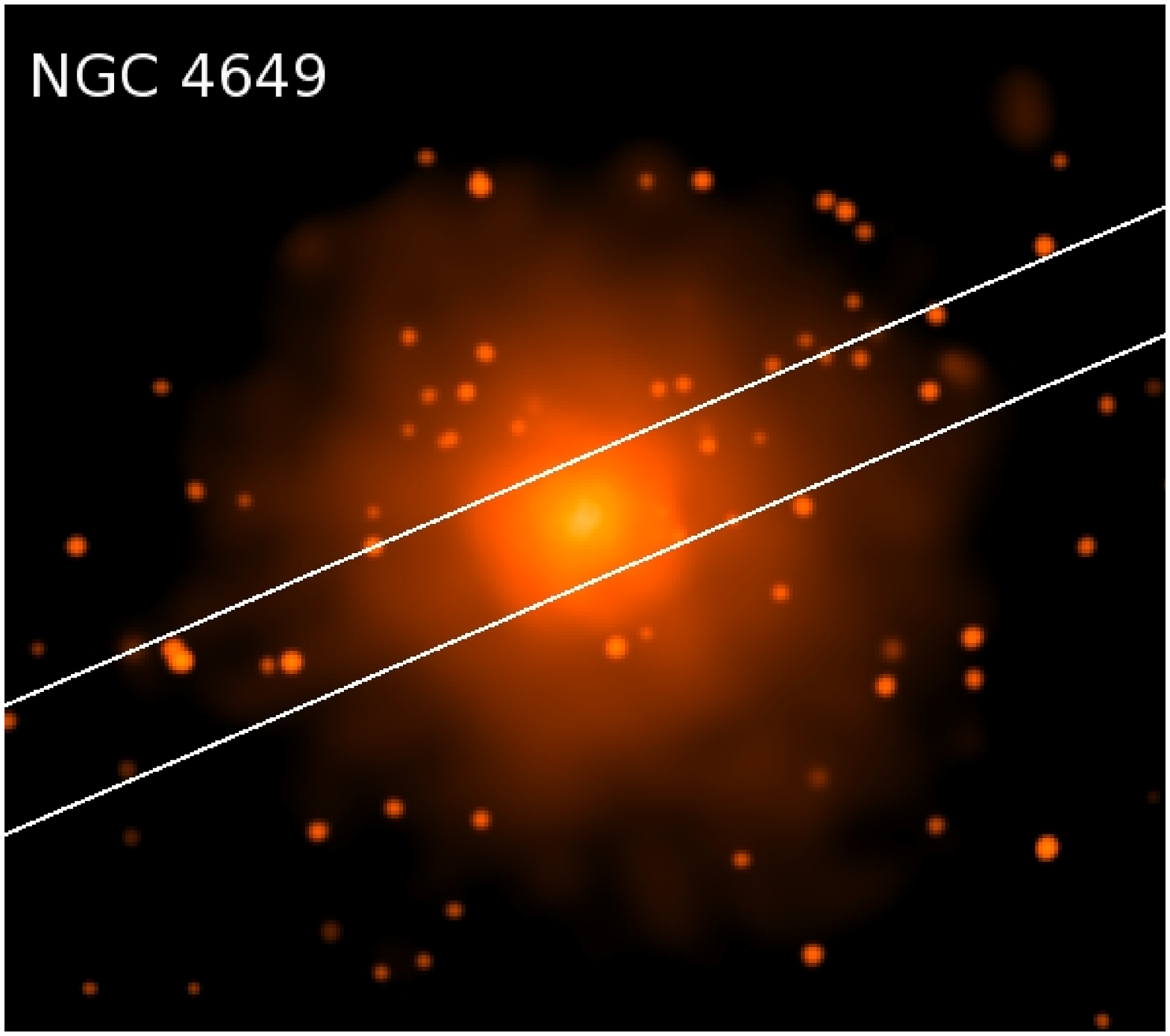}
\end{minipage}
\begin{minipage}{0.48\textwidth}
\includegraphics[width=0.85\textwidth,clip=t,angle=0.]{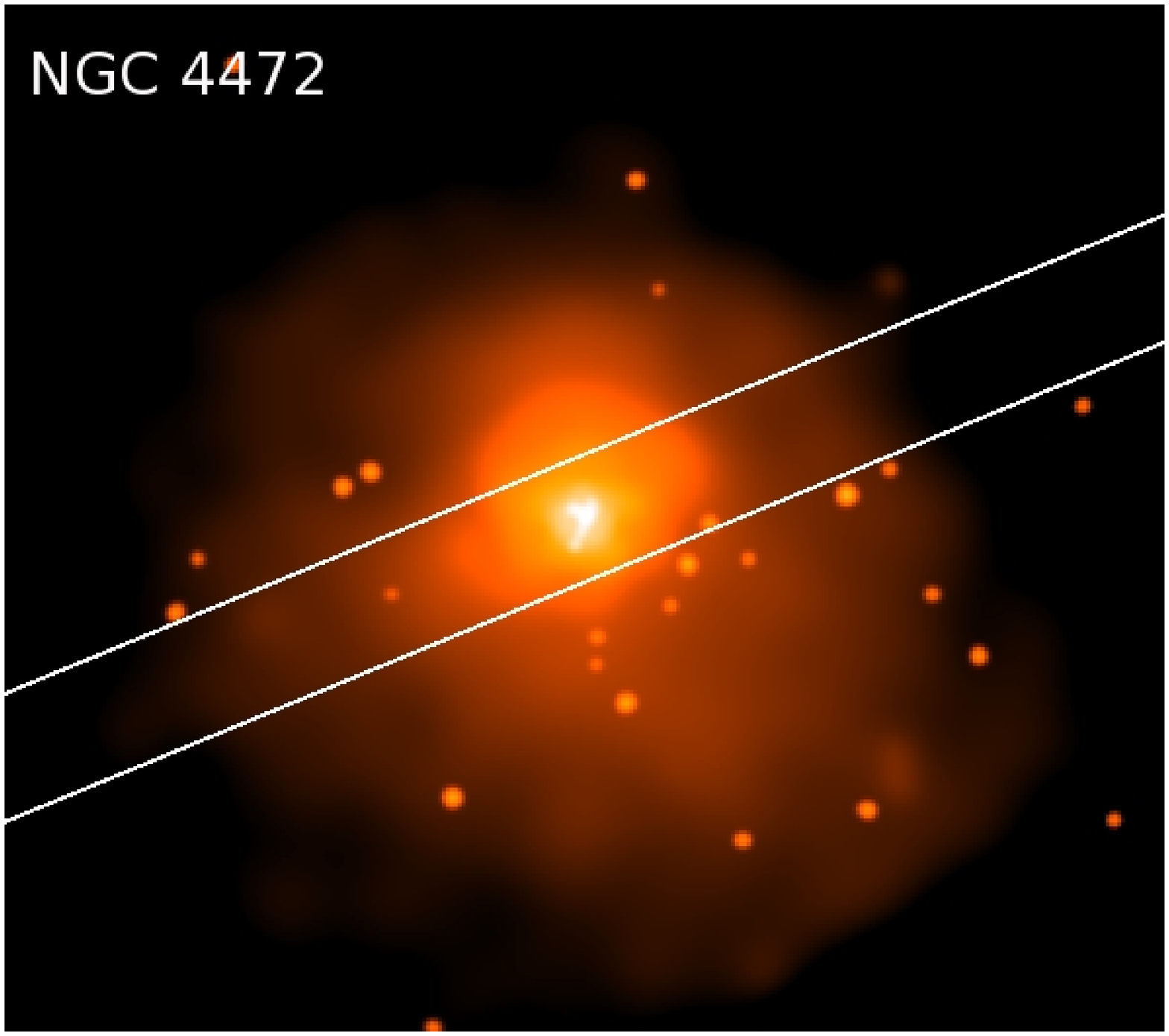}
\end{minipage}
\caption{{\it{Chandra}} images of the giant elliptical galaxies in our sample with over-plotted RGS extraction regions (see also Fig.~\ref{fig:ngc4636}). The extraction regions are 0.5\arcmin\ wide. While the core of NGC~5813 is strongly disturbed, the cores of the other three galaxies are relatively relaxed. The images were extracted in the 0.5--2.0 keV band, cleaned and adaptively smoothed. } 
\label{fig:Chandrasample}
\end{figure*}

We have analyzed a sample of five nearby, giant elliptical galaxies observed with \xmm\ RGS. The sample is listed in Table~\ref{obslog}. The galaxies were selected based on their relatively low core temperatures, at which a significant fraction of Fe is expected to be in the form of \ion{Fe}{xvii} (see Fig.~\ref{fractions}), their large X-ray fluxes, and strongly peaked surface brightness distributions, which are necessary to obtain RGS spectra with sufficient statistics. The sample contains both relatively relaxed galaxies and systems showing very clear signs of interaction between the hot halo gas and central Active Galactic Nucleus (AGN). In Figs.~\ref{fig:ngc4636} and \ref{fig:Chandrasample} we show 0.5--2.0~keV {\it{Chandra}} images of the galaxies in our sample with the RGS extraction regions overplotted. The images show that the cores of NGC~4649, NGC~4472, and NGC~1404 are {\it{relatively}} relaxed \citep[although on larger scales NGC~1404 shows a prominent cold front to the Northwest, and NGC~4649 and NGC~4472 harbor central radio sources, which exhibit interaction with the hot gas,][]{shurkin2008,biller2004}. NGC~4636 and NGC~5813, on the other hand, look much more disturbed, with obvious signs of recent episodes of interaction between the hot plasma and the AGN. NGC~4636 has a very dense core and X-ray bright arm-like structures which might have been produced by shocks \citep{jones2002}. The core of NGC~5813 is highly disturbed with most of the hot gas being pushed into dense sheets by the rising bubbles of relativistic plasma. 

Currently the best data set for the study of resonant scattering is that for NGC~4636. The X-ray halo of this galaxy has a conveniently low temperature, at which $\sim$26\% of Fe is in the form of \ion{Fe}{xvii}, a very dense core and a deep \xmm\ observation, which also allows us to extract spectra from outside the core region with reasonably good statistics in the \ion{Fe}{xvii} lines. Using a deep \chandra\ observation of this galaxy we derive deprojected density and temperature profiles which we use to model radial intensity profiles for the strongest resonance lines, taking into account the effects of resonant scattering for different assumed values of turbulent velocities.

\section{Data analysis}

\begin{table*}
\begin{center}
\caption{The best fit parameters for a single-temperature, optically thin plasma model fitted to the \xmm\ RGS spectra extracted from 0.5\arcmin\ wide regions centred on the cores of the galaxies. For NGC~4636 we also show the results from fits to spectra extracted from two 2.25\arcmin\ wide regions surrounding the core (see Fig.~\ref{fig:ngc4636}). The 13.8--15.5~\AA\ part of the spectrum, where the strongest \ion{Fe}{xvii} and \ion{Fe}{xviii} resonance lines are present was initially excluded from the fits. Fluxes are given in the 0.3--2.0~keV band. The emission measure is defined as $Y = \int n_{\mathrm{e}} n_{\mathrm{H}} dV$. The scale factor $s$ is the ratio of the observed LSF width to the expected LSF for a flat abundance distribution. Abundances are quoted with respect to the proto-solar values of \citet{lodders2003}. The last three rows list the best fit line ratios in the full spectral band (after the \ion{Fe}{xvii} ion was set to zero in the model and replaced by gaussian lines), the theoretical line ratios predicted for an optically thin plasma, and the derived level of suppression of the 15.01~\AA\ line, $(I/I_0)_{15.01\AA}$. }
\begin{tabular}{lcccccc}
\hline\hline
galaxy                               									&  NGC 4636 core  & NGC 4636 outer reg. 		&  NGC 5813 			&   NGC 1404			&   NGC 4649			&   NGC 4472   	\\
\hline
flux~($10^{-12}$~erg~cm$^{-2}$)							& $1.75\pm0.08$	& $2.57\pm0.17$	& $1.47\pm0.12$		&  $1.08\pm0.11$ 		&  $1.63\pm0.10$		& $1.44\pm0.08$		\\
$Y$~($10^{64}$~cm$^{-3}$)								& $0.47\pm0.02$	& $0.46\pm0.03$ 	& $1.01\pm0.10$		& $0.50\pm0.05$		&  $0.31\pm0.03$		& $0.34\pm0.02$		\\
$kT$ (keV)                           									& $0.606\pm0.006$  & $0.695\pm0.004$	& $0.645\pm0.008$  	&  $0.608\pm0.009$  	&  $0.774\pm0.007$     	& $0.781\pm0.006$  	\\
$s$                              										& $0.40\pm0.04$ 	& $1.02\pm0.04$	&  $0.87\pm0.11$		&   $0.97\pm0.22$		&  $0.69\pm0.21$  		& $0.79\pm0.12$         	\\
N  													&  $1.3\pm0.3$  	& $1.5\pm0.4$ 		&  $2.0\pm0.8$  		&  $2.3\pm0.8$			&  $1.3\pm0.7$ 		& $1.3\pm0.5$ 			\\
O 													& $0.44\pm0.05$ 	& $0.61\pm0.06$	&  $0.53\pm0.09$        	&  $0.58\pm0.10$		&  $0.61\pm0.15$		& $0.53\pm0.07$		\\
Ne 													& $0.31\pm0.08$	& $0.39\pm0.18$	&  $0.33\pm0.19$		& $0.81\pm0.22$		&  $1.31\pm0.35$ 		& $1.18\pm0.22$ 		\\
Fe  													& $0.52\pm0.03$	& $0.92\pm0.06$	&  $0.75\pm0.09$		& $0.67\pm0.08$		&  $0.87\pm0.18$		& $0.83\pm0.08$ 		\\
$\lbrack(I_{\lambda17.05}+I_{\lambda17.10})/I_{\lambda15.01}\rbrack_{\mathrm{observed}}$ 			& $2.04\pm0.21$ 		&$1.28\pm0.13$	&  $1.99\pm0.34$ 		& $1.98\pm0.29$		&  $1.25\pm0.28$		& $2.24\pm0.34$ 		\\
$\lbrack(I_{\lambda17.05}+I_{\lambda17.10})/I_{\lambda15.01}\rbrack_{\mathrm{predicted}}$			& 1.31	&    1.31			& 1.30				& 1.31				&   1.27				&  1.27  \\
$I_{15.01}/I_{\mathrm{0}~15.01}$							& $0.64\pm0.07$		& $1.02\pm0.10$ &    $0.65\pm0.11$		& $0.66\pm0.10$		&    $1.02\pm0.23$ 				& $0.57\pm0.09$		\\
\hline
\label{main}
\end{tabular}
\end{center}
\end{table*}

\begin{figure*}
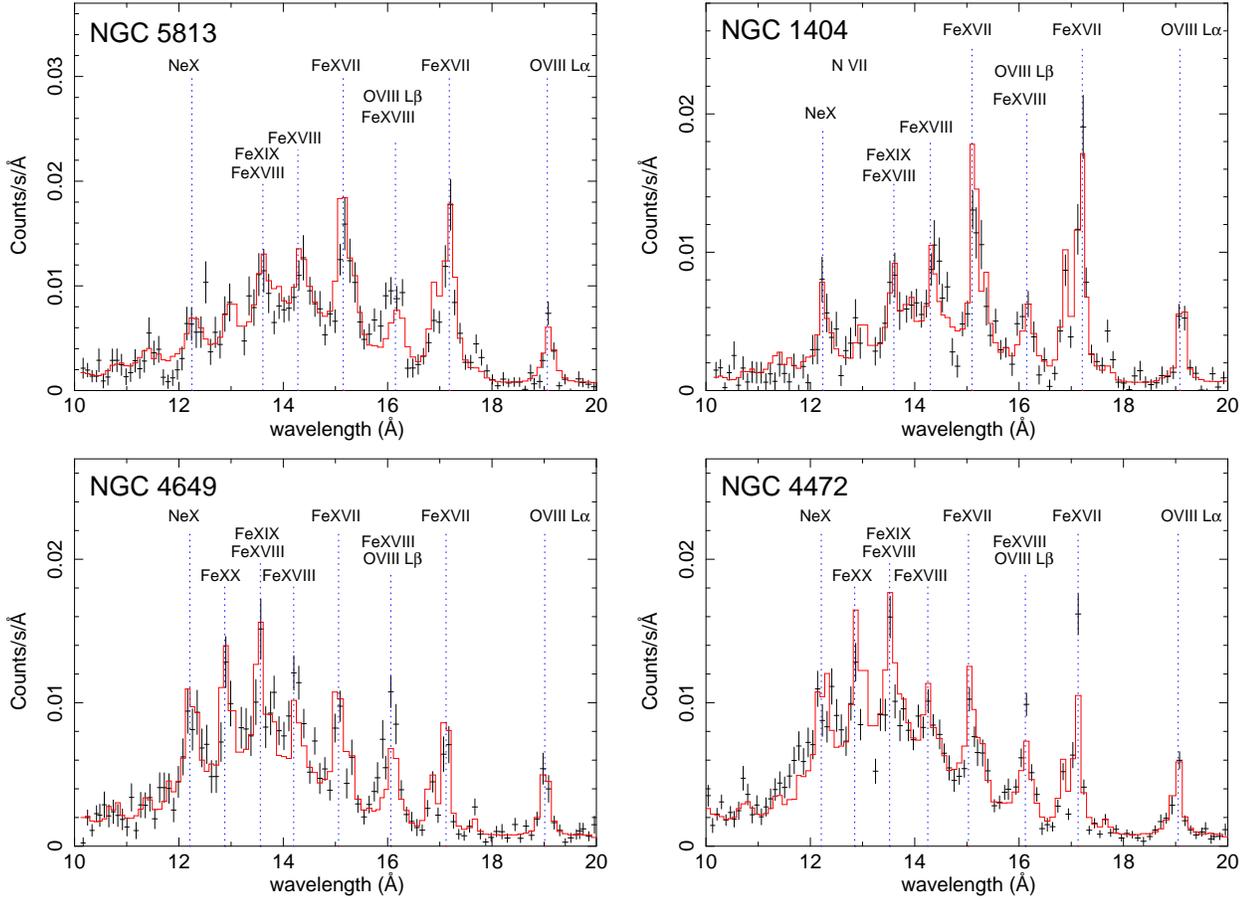

\begin{minipage}{0.47\textwidth}
\includegraphics[width=0.70\textwidth,clip=t,angle=270.]{NGC5813narrow.ps}
\end{minipage}
\begin{minipage}{0.47\textwidth}
\includegraphics[width=0.70\textwidth,clip=t,angle=270.]{NGC1404narrow.ps}
\end{minipage}\\
\vspace{2mm}

\begin{minipage}{0.47\textwidth}
\includegraphics[width=0.70\textwidth,clip=t,angle=270.]{NGC4649narrow.ps}
\end{minipage}
\begin{minipage}{0.47\textwidth}
\includegraphics[width=0.70\textwidth,clip=t,angle=270.]{NGC4472narrow.ps}
\end{minipage}
\caption{RGS spectra for NGC~5813, NGC~1404, NGC~4649, and NGC~4472. The spectra were extracted from 0.5\arcmin\ wide extraction regions centred on the cores of the galaxies. The full line indicates the best-fit  optically-thin single-temperature plasma model. The 13.8--15.5~\AA\ part of the spectrum, where the strongest \ion{Fe}{xvii} and \ion{Fe}{xviii} resonance lines are present, was excluded from these spectral fits.} 
\label{fig:RGSsample}
\end{figure*}

\subsection{\xmm\ RGS data analysis}

We processed the RGS data using version 8.0.0 of the \xmm\ Science Analysis System (SAS), and extracted spectra using the method described by \citet{tamura2001}. To minimize the contamination by soft-protons from Solar flares, we extracted a light curve for each dataset using events on CCD~9 of the RGS, outside the central region, with a distance larger than 30\arcsec\ from the dispersion axis, and excluded time intervals with elevated count rates. The original exposure time and the total time after cleaning are listed in Table~\ref{obslog}. The flares in the observation of NGC~1404 were weak and we obtain the same best fit spectral parameters for both the filtered and unfiltered data set. Therefore, for this galaxy we used the full observation. 
Because emission from the galaxies fills the entire field of view of the RGS, the background cannot be estimated from a region on the detector away from the source. Therefore, we modeled the background using the standard RGS background model for extended sources available in SAS \citep[rgsbkgmodel,][]{riestra2004}.

The effects of resonant scattering are most significant in the centermost parts of the galaxies, where the density is highest. Therefore, we extract spectra from narrow 30\arcsec\ wide bands centred on the cores of the galaxies, probing regions within radii of 1.3--2.2~kpc (depending on distance) in the cross dispersion direction of the RGS. We extract and fit both first and second order spectra. 
Because the RGS operates without a slit, it collects all photons from within the 0.5\arcmin$\times\sim$12\arcmin\ field of view. Line photons originating at angle $\Delta\theta$ (in arcminutes) along the dispersion direction, will be shifted in wavelength by 
\begin{eqnarray}
\Delta\lambda=0.138\, \Delta\theta~\AA. 
\end{eqnarray}
Therefore, every line will be broadened by the spatial extent of the source.
To account for this broadening in our spectral model, for each RGS spectrum, we produce a predicted line spread function (LSF) by convolving the RGS response with the surface brightness profile of the galaxy derived from the EPIC/MOS1 image in the 0.8--1.4~keV band along the dispersion direction. 
Because the radial profile of a particular spectral line can be different from the overall radial surface brightness profile (e.g. in the presence of metallicity gradients), the line profile is multiplied by a scale factor $s$, which is the ratio of the observed LSF width to the expected LSF for a flat abundance distribution. This scale factor is a free parameter in the spectral fit.

For the spectral modeling of the RGS data we use the SPEX package \citep{kaastra1996}. Spectral fitting of the RGS data is done in the 10~\AA\ to 28~\AA\ band.  We model the emission of the observed galaxies as optically thin plasmas in collisional ionization equilibrium, absorbed by neutral Galactic gas with Solar abundances. For each object we fix the Galactic column density to the value determined by the Leiden/Argenitine/Bonn (LAB) Survey of Galactic \ion{H}{i} \citep[][]{kalberla2005}. For the emission of the elliptical galaxies we assume a single-temperature plasma. Using more complicated multi-temperature models does not improve the fits and differential emission measure models always converge to a simple single-temperature approximation. The spectral normalization, plasma temperature, and the abundances of N, O, Ne, and Fe are free parameters in the fit. For the spectral fitting we use C-statistics. 

\subsection{\chandra\ analysis of NGC~4636}
\label{chandradata}

We reprocessed the \chandra\ observations of NGC~4636 (OBSID 3926 and 4415) applying the latest CTI and time-dependent gain calibrations using standard methods \citep[see][for more details]{vikhlinin2005}. In brief, we performed the usual filtering by grade, excluded bad/hot pixels and columns, removed cosmic ray `afterglows', and applied the VF mode filtering. We excluded data with anomalously high background. The remaining exposure time for NGC~4636 is 150.4~ks. The background files were processed in the same manner as the observations.  An additional background component is produced during the 41~ms of the 3.2~s nominal exposure readout of the ACIS CCDs, while the chips are exposed to the sky.  This small contribution of the source flux, 1.3\%, is uniformly re-distributed along the readout direction and is subtracted using the technique
described by \citet{markevitch2000}.  The blank-field background is also renormalized by reducing its integration time by 1.3\% to account for this additional subtraction. 

As the final step in data preparation \citep[see][for more details]{churazov2008}, we computed, for each X-ray event, the ratio of the effective area (a function of photon energy and position) to that at a predefined energy and position: 
\begin{eqnarray}
\eta=A(E,x_d,y_d)/A_0(E),
\label{eqn:eff}
\end{eqnarray}
where $A(E,x_d,y_d)$ includes mirror and detector efficiencies (including non-uniformity of the detector quantum efficiency and the time and spatially dependent contamination on the optical blocking filter).  Finally, for each event list, we make an exposure map that accounts for all position dependent, but energy independent,
efficiency variations across the focal plane (e.g., overall chip geometry, dead pixels or rows, and variation of telescope pointing direction). We use these data in Sect.~\ref{modelling}  to derive deprojected gas density and
temperature profiles.

\section{Observations of resonant scattering} 

\begin{figure*}
\begin{minipage}{0.47\textwidth}
\includegraphics[width=0.70\textwidth,clip=t,angle=270.]{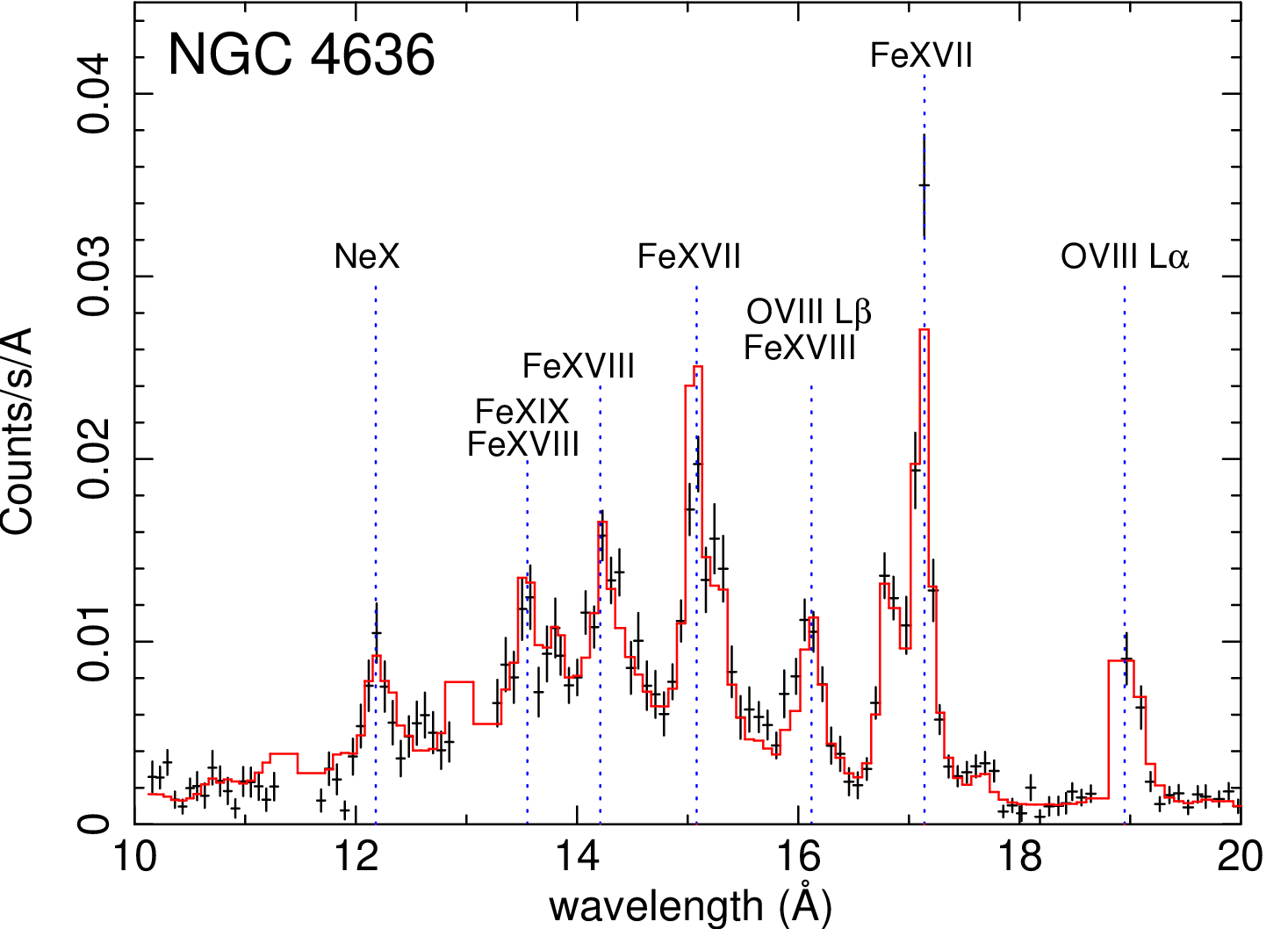}
\end{minipage}
\begin{minipage}{0.47\textwidth}
\includegraphics[width=0.70\textwidth,clip=t,angle=270.]{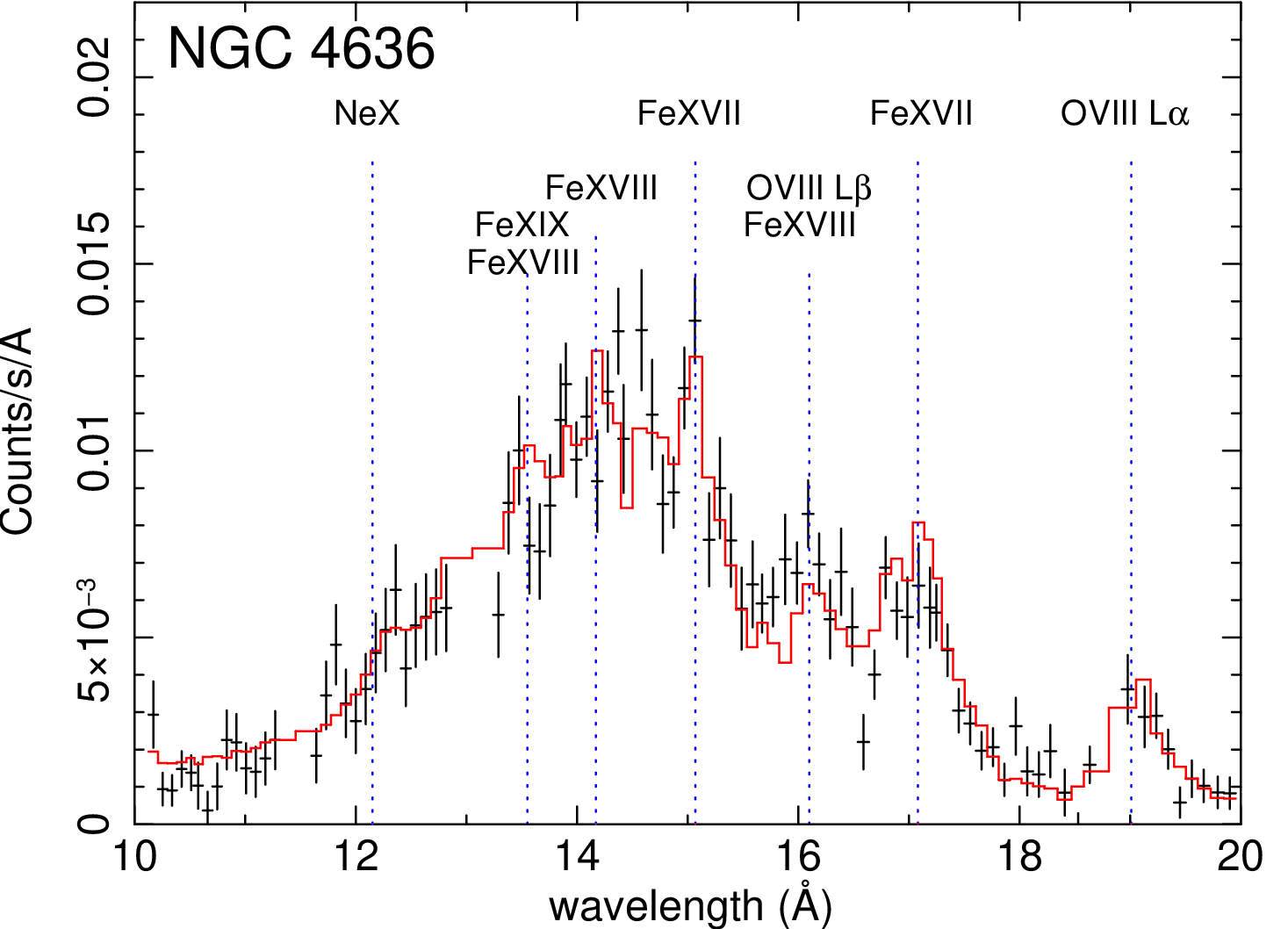}
\end{minipage}\\
\caption{\xmm\ RGS spectra extracted from a 0.5\arcmin\ wide extraction region centered on the core of NGC~4636 (left panel) and from two 2.25\arcmin\ wide extraction regions surrounding the core (right panel). The ratios of the \ion{Fe}{xvii} lines at 15.01 and $\sim$17~\AA\ change. The full line indicates the best fit  optically thin single-temperature plasma model. The 13.8--15.5~\AA\ part of the spectrum, where the strongest \ion{Fe}{xvii} and \ion{Fe}{xviii} resonance lines are present, was excluded from these spectral fits.} 
\label{fig:NGC4636}
\end{figure*}

The results from the spectral fitting of the \xmm\ RGS data with an optically thin single-temperature plasma model are shown in Table~\ref{main} and Figures \ref{fig:RGSsample} and \ref{fig:NGC4636}. The 13.8--15.5~\AA\ part of the spectrum, where the strongest \ion{Fe}{xvii} and \ion{Fe}{xviii} resonance lines are present, was initially excluded from this fit. These lines are expected to be suppressed by resonant scattering and this suppression, which is not accounted for in the plasma model, could slightly bias the best fit spectral parameters. After obtaining the best fit, we freeze the temperature and Fe abundance of the thermal component, set the abundance of the \ion{Fe}{xvii} ion to zero in the model, add narrow Gaussians at the wavelength of the strongest \ion{Fe}{xvii} lines (restframe wavelengths of 15.01~\AA, 17.077~\AA, 15.30~\AA, and 16.80~\AA) and re-fit the full 10--28~\AA\ spectral band. Freezing the temperature and the Fe abundance is necessary because by setting the \ion{Fe}{xvii} ion to zero and replacing it with Gaussians in the model we lose important constraints on these parameters. The line normalisations, the continuum level (normalisation of the thermal component), the abundances of the other elements and the broadening factor $s$ are free parameters. In the error calculation, we also marginalise over the uncertainties of the temperature and Fe abundance as determined in the initial fit. We determine the best fitting normalisations for these Gaussians and the value of $(I_{\lambda17.05}+I_{\lambda17.10})/I_{\lambda15.01}$ (the 17.05 and 17.10 \AA\ lines are blended and we fit them with a single Gaussian at 17.077 \AA). We divide the theoretical ratio $(I_{\lambda17.05}+I_{\lambda17.10})/I_{\lambda15.01}$ by the measured value, which, assuming the lines at $\sim$17~\AA\ are optically thin, gives us the level at which the 15.01~\AA\ line is suppressed,  
$(I/I_0)_{15.01\AA}$. These values are listed in Table~\ref{main}. In four out of five galaxies the line ratios are significantly higher (observed values $\sim2.0$) than the expected theoretical value for an optically thin plasma of 1.27--1.31, indicating that the 15.01~\AA\ line is optically thick. In NGC~4649, the line ratios are consistent with the 15.01~\AA\ line being optically thin. 

A number of factors argue that the observed high $(I_{\lambda17.05}+I_{\lambda17.10})/I_{\lambda15.01}$ line ratio is due to resonant scattering. 
The  $(I_{\lambda17.05}+I_{\lambda17.10})/I_{\lambda15.01}$ line ratio in an optically thin plasma can reach values $\sim1.9$ only for temperatures as low as 0.2 keV \citep[see Fig.~\ref{fractions},][]{doron2002}, which is much lower than the observed temperatures of the galaxies. Moreover, if the plasma in the cores of the galaxies were multiphase with an additional 0.2 keV component, which would alter the \ion{Fe}{xvii} line ratios, then we would also observe other strong lines (e.g. \ion{O}{vii}), which we do not. Another possibility to obtain such high line ratios in an optically thin plasma is non-equilibrium ionization (NEI). However, NEI effects are unlikely in the dense plasma in cores of giant elliptical galaxies because, due to the high density, the equilibration time is short. If the higher-than-expected line ratios were due to incomplete atomic physics in the current spectral models, then they should not change as a function of radius. This can be directly verified in NGC~4636, which has the best combination of X-ray flux, temperature, and exposure. The data for this galaxy allow us to extract spectra and determine the $(I_{\lambda17.05}+I_{\lambda17.10})/I_{\lambda15.01}$  line ratios in the combined spectrum from two regions surrounding the core at 15\arcsec--150\arcsec\ from the centre of the galaxy, in the crossdispersion direction of the RGS (see Fig.~\ref{fig:ngc4636}). We follow here the same fitting procedure as for the cores of the galaxies. The best fitting results are shown in the third column of Table~\ref{main}. The off-centre line ratios are consistent with the 15.01~\AA\ line being optically thin in this region. The difference in the spectra of the core and surrounding regions of NGC~4636 can be seen clearly in Fig.~\ref{fig:NGC4636}. While the 15~\AA\ line is suppressed in the centre, it may be slightly enhanced outside the dense core.

\begin{table*}
\begin{center}
\caption{Oscillator strengths and optical depths for the strongest X-ray lines in the spectrum of NGC~4636 for Mach numbers 0.0, 0.25, 0.5, 0.75. The expected suppression due to resonant scattering  $(I/I_0)_{\mathrm{circ}}$ gives the value integrated within a radius of 2~kpc from the centre of the galaxy. The approximate values $(I/I_0)_{\mathrm{RGS}}$ were obtained by integrating within an ``effective extraction region'' which is 0.5\arcmin\ wide and 3\arcmin\ long. The metallicity is assumed to be constant with the radius. }
\label{table:depths}
\begin {tabular}{cccc@{}c@{}cc@{}c@{}cc@{}c@{}cc@{}c@{}c}
\hline
\hline
Ion 					& $\lambda$ (\AA) 	& $f$ 	& \multicolumn{3}{c}{$M=0.0$} 	& \multicolumn{3}{c}{$M=0.25$} 	& \multicolumn{3}{c}{$M=0.5$} & \multicolumn{3}{c}{$M=0.75$} \\
					&				&		&  $\tau$ 	& $(I/I_0)_{\mathrm{circ}}$ & $(I/I_0)_{\mathrm{RGS}}$ & $\tau$ 	& $(I/I_0)_{\mathrm{circ}}$  & $(I/I_0)_{\mathrm{RGS}}$& $\tau$  &  $(I/I_0)_{\mathrm{circ}}$ & $(I/I_0)_{\mathrm{RGS}}$ & $\tau$ & $(I/I_0)_{\mathrm{circ}}$ & $(I/I_0)_{\mathrm{RGS}}$ \\
\hline
\ion{Fe}{xvii} 			&  15.01 			& 2.73 	& 8.8 	& 0.47 & 0.69		& 3.6 	& 0.59 & 0.79 			& 1.9		& 0.68	& 0.85 & 1.3   & 0.76 & 0.89   	\\
\ion{Fe}{xvii} 			&  17.05 			& 0.12 	& 0.5 	& 0.89 & 0.95		& 0.2 	& 0.95 & 0.98			& 0.1		& 0.97  	& 0.99 & 0.07 & 0.98 & 0.99	\\
\ion{Fe}{xviii} 			&  14.20 			& 0.57 	& 1.3 	& 0.72 & 0.87		& 0.5		& 0.86 & 0.94	 		& 0.3		& 0.91	& 0.96 & 0.2 & 0.95  & 0.98	 \\
\ion{Fe}{xviii}			& 16.08 			& 0.005 	& 0.01	& 0.99 & 1.00		& 0.005     & 0.99 & 1.00			& 0.003  	& 1.00	& 1.00 & 0.002& 1.00 & 1.00	 \\
\ion{O}{viii}~L$\alpha$	& 18.96 			& 0.28 	& 1.2 	& 0.74 & 0.87		& 0.8 	& 0.81 & 0.91			& 0.5  	& 0.88	& 0.94 & 0.3 & 0.91 & 0.96	\\
\hline
\end {tabular} 
\end{center}
\end{table*}

The best fit line broadening scale factor $s$ indicates that in NGC~4636 the distribution of the line producing ions is more peaked than the surface brightness, suggesting that the distribution of metals is centrally peaked.  In the other systems this effect is smaller and less significant.

\section{Modelling of resonant scattering in NGC~4636}
\label{modelling}

\begin{figure}
\includegraphics[width=1.0\columnwidth,clip=t,angle=0.]{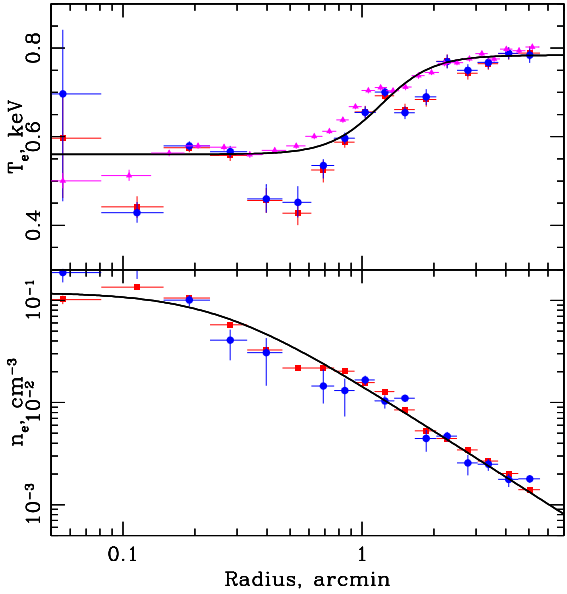}
\caption{Observed radial profiles of the electron density and temperature used to model resonant scattering in NGC~4636. The blue circles and red squares indicate the deprojected profiles determined from \chandra\ data with metallicity as a free parameter in the fit and with metallicity fixed to 0.68 Solar, respectively. The magenta triangles show the projected radial temperature profile with the metallicity as a free parameter in the fit. 
The solid lines show the parametrization of the gas density and gas temperature profiles given in eqs.~\ref{ne_prof} and \ref{kt_prof}.}
\label{input}
\end {figure}

To gain a better understanding of the observed line ratios, we simulate the effects of resonant scattering on the radial intensity profiles of the strongest X-ray emission lines in NGC~4636. For this we use the deprojected temperature and density profiles measured with \chandra. 

In our deprojection analysis, we follow the approach described by \citet{churazov2008}. We assume spherical symmetry, but make no specific assumption about the form of the underlying gravitational potential. For a given surface brightness profile in $n_\mathrm{a}$ annuli, we choose a set of $n_\mathrm{s}$ ($n_\mathrm{s}\le n_\mathrm{a}$) spherical shells with the inner radii $r(\mathrm{i}),~\mathrm{i}=1,\ldots,n_\mathrm{s}$. The gas emissivity $\mathcal{E}$ is assumed to be uniform inside each shell, except for the outermost shell, where the gas emissivity is assumed to decline as a power law of radius: $\mathcal{E}= \mathcal{E}_{\mathrm{out}} r^{-6\beta_{\mathrm{out}}}$, where $\beta_{\mathrm{out}}$ is a parameter.  The deprojection process is a simple least square solution to determine the set of emissivities in the set of shells (along with the emissivity normalization $\mathcal{E}_{\mathrm{out}}$ of the outer layers) that provides the best description of the observed surface brightness. The emissivity of each shell can then be evaluated as an explicit linear combination of the observed quantities.  Since the whole procedure is linear, the errors in the observed quantities can be propagated straightforwardly. With our definition of $\eta$ (see subsection \ref{chandradata}), the projection matrix does not depend on energy. Therefore, the deprojection in {\em any energy band} is precisely that used to deproject the surface brightness.  Thus, we can accumulate a set of spectra (corrected for background and readout) for each of the $n_\mathrm{a}$ annuli, and apply the deprojection to determine the emissivities of each shell in each of the ACIS energy channels.  

The \chandra\ data were modeled using XSPEC V12 \citep{arnaud1996} and the APEC model \citep{smith2001}. The gas temperature and normalization were free parameters in the model. The heavy element abundance was either a free parameter in the fit or fixed to 0.68 Solar (as determined from a spectral fit to \chandra\ data extracted within the radius of 1\arcmin\ from the centre of NGC~4636). The metallicity profile in NGC~4636 has large uncertainties, because the metal abundance distribution in the deprojected \chandra\ spectra has a large scatter as a result of noise enhancement during deprojection coupled with the limited spectral resolution of the CCDs.

The electron density profile was derived from the spectral normalization, fixing the proton to electron ratio to $0.83$.  Fig. \ref{input} shows the deprojected electron density (lower panel) and the temperature (upper panel) as a function of distance from the centre of the galaxy. The circles and squares show the deprojected profiles for metallicity as a free parameter and for metallicity fixed to 0.68 Solar, respectively. The triangles show the projected temperature profile with the metallicity as a free parameter. 

Based on the observed radial distributions of electron density $n_{\mathrm{e}}$ and temperature $kT_{\mathrm{e}}$ in NGC~4636, we adopt the following approximate forms for the deprojected 
density and temperature profiles:

\begin{equation}
n_{\mathrm{e}}=1.2\times10^{-1} \left[1+\left(\frac{r}{0.25'}\right)^2\right]^{-0.75}~\mathrm{cm}^{-3},
\label{ne_prof}
\end{equation}

and

\begin{equation}
kT=0.56\frac{1+1.4(r/1.2')^4}{1+(r/1.2')^4}~\mathrm{keV.}
\label{kt_prof}
\end{equation}

We consider two possible radial behaviors for metallicity: a simple constant abundance of 0.68 Solar; and a centrally peaked abundance distribution, which is more consistent with the best fit \xmm\ RGS line profile ($s=0.40\pm0.04$).  In the latter case we approximate the metallicity profile with the following function:
\begin{equation}
Z=0.95\frac{2+(r/0.8')^3}{1+(r/0.8')^3}-0.4. 
\end{equation}

We calculate the optical depth of the 15~\AA\ line from the centre of the galaxy to infinity, $\tau=\int n_{\mathrm{i}}\sigma_{\mathrm{0}} \mathrm{d} r$, where $n_\mathrm{i}$ is the ion concentration and $\sigma_{\mathrm{0}}$ is the cross section at the line centre, which for a given ion is

\begin{equation}
\sigma_0=\frac{\sqrt{\pi}hr_{\mathrm{e}} cf}{\Delta E_\mathrm{D}},
\end{equation}
where the Doppler width is given by
\begin{equation}
\Delta E_\mathrm{D}=E_0\left(\frac{2kT_\mathrm{e}}{Am_\mathrm{p} c^2}+\frac{V_\mathrm{turb}
  ^2}{c^2}\right)^{1/2}.
\end{equation}
In these equations $r_\mathrm{e}$ is the classical electron radius, $f$ is the oscillator strength of a given atomic transition, $E_0$ is the rest energy of a given line, $A$ is the atomic mass of the corresponding element, 
$m_{\mathrm{p}}$ is the proton mass, $c$ is the speed of light, and $V_\mathrm{turb}$ is the characteristic velocity of isotropic turbulence. $V_\mathrm{turb}\equiv \sqrt{2} V_\mathrm{1D, turb}$, where $V_\mathrm{1D, turb}$ is the velocity dispersion in the line of sight due to turbulence. 
Using the adiabatic sound speed, $c_{\mathrm{s}}=\sqrt{\gamma\, k\, T\, /\mu\, m_{\mathrm{p}}}$, the expression for the broadening can be rewritten as
\begin{equation}
\Delta E_\mathrm{D}=E_0\left[\frac{2kT_\mathrm{e}}{Am_\mathrm{p} c^2}(1+\frac{\gamma}{2\, \mu} AM^2)\right]^{1/2}, 
\end{equation}
where $\mu=0.6$ is the mean particle mass, $\gamma$ is the adiabatic index, which for ideal monatomic gas is 5/3, and $M=V_{\mathrm{turb}}/c_{\mathrm{s}}$ is the corresponding Mach number.  The line energies and oscillator strengths were taken from ATOMDB\footnote{http://cxc.harvard.edu/atomdb/WebGUIDE/index.html} and the NIST Atomic Spectra Database\footnote{www.physics.nist.gov/PhysRefData/ASD/index.html}.

\begin{figure}
\includegraphics[width=0.5\textwidth,clip=t,angle=0.]{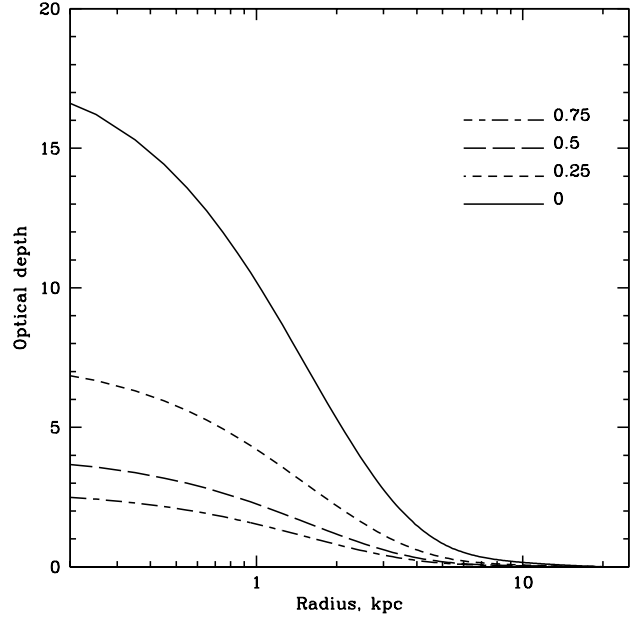}
\caption{Optical depth of the 15.01~\AA\ line calculated from the given radius to infinity, for isotropic turbulent velocities corresponding to Mach numbers 0.0, 0.25,
0.5, and 0.75. A centrally peaked Fe abundance distribution is assumed.} 
\label{depth15Amodel}
\end{figure}

\begin{figure}
\includegraphics[width=0.5\textwidth,clip=t,angle=0.]{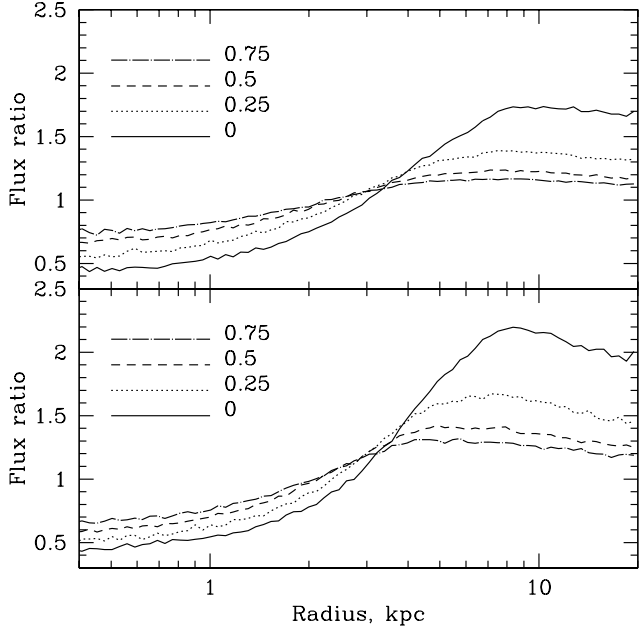}
\caption{Radial profiles of the ratio of the 15.01~\AA\ line intensity with and without the effect of resonant scattering, for isotropic turbulent velocities corresponding to Mach numbers 0.0, 0.25, 0.5, and 0.75. The upper panel shows the profiles for a constant metallicity, and the lower panel for a centrally peaked Fe abundance distribution.} 
\label{flux15Amodel}
\end{figure}

The resonant scattering effect has been calculated using Monte-Carlo simulations, as described by \citet{churazov2004}. We model the hot halo as a set of spherical shells and obtain line emissivities for each shell using the APEC plasma model \citep{smith2001}. We account for scattering by assuming a complete energy
redistribution and dipole scattering phase matrix. The line of \ion{Fe}{xvii} has a pure dipole scattering phase matrix. In Fig.~\ref{depth15Amodel} we show the optical depth of the 15.01~\AA\ \ion{Fe}{xvii} line as a function of radius, for various turbulent velocities, assuming the peaked Fe abundance distribution. For higher turbulent velocities, the optical depth becomes smaller.
In Fig.~\ref{flux15Amodel} we show the ratio of the 15.01~\AA\ line intensity calculated including the effect of resonant scattering to the line intensity without accounting for resonant scattering for various turbulent velocities, both for constant and peaked iron abundance distributions. The figure clearly shows the effect of the resonant scattering: in the core of the galaxy the intensity of the line is suppressed, whereas in the surrounding regions the intensity of the line rises. The effect is more prominent in the case of the centrally peaked Fe abundance distribution, where the central suppression is larger and the enhancement in the outer part is stronger. 

We compare the models of resonant scattering with the observations and determine the systematic uncertainties involved by simulating spectra corresponding to each of these models, which are then fit in the same way as the observed data. To this end, we multiply the surface brightness profile of NGC~4636 with the theoretical suppression profiles for the 15.01~\AA\ line (Fig. \ref{flux15Amodel}) to determine the predicted RGS line profile for the four considered characteristic turbulent velocities, for both assumed abundance distributions. We simulate spectra with photon statistics comparable to that of the actual observation, using the best fit values for the core of NGC~4636 from Table~\ref{main} as input parameters, and convolving the 15.01~\AA\ line through these predicted line profiles. 
By fitting the data simulated with no turbulence, we obtain $(I/I_{0})_{15.01\AA}=0.69$ and 0.71 for peaked and flat abundance profiles, respectively. 
The ratios of $(I/I_0)_{15.01\AA}$ for isotropic turbulent velocity of $M=0.25$ are 0.76 and 0.78, and for $M=0.50$ they are 0.80 and 0.82, for peaked and flat abundance profiles, respectively. 
The systematic uncertainty on these values due to our lack of knowledge about the actual Fe abundance distribution is $\pm$0.02. We add this uncertainty in quadrature to the uncertainties associated with our modeling of the line broadening, which is at most 0.03, resulting in a total uncertainty of 0.04. Thus we conclude that the turbulent velocities in NGC~4636 are relatively small and that isotropic turbulence with a characteristic velocity of $M>0.25$ can be ruled out at the 90\% confidence level; turbulent gas motions at $M>0.5$ can be ruled out at the 95\% confidence level. (The sound speed in NGC~4636 is $c_{\mathrm{s}}\sim400$~km~s$^{-1}$.) 

In Table~\ref{table:depths}, we list the strongest lines observed in the spectrum of NGC~4636, their oscillator strengths and
optical depths for different Mach numbers. Clearly, \ion{Fe}{xvii} at 15.01~\AA\ is the most optically thick line in the spectrum, but \ion{O}{viii}~L$\alpha$ and \ion{Fe}{xviii} at 14.2~\AA\ also have relatively large optical depths. For $M=0$, the depth of the \ion{Fe}{xvii} line at 17.05~\AA\ line is $\tau=0.5$, meaning that our assumption that the 17.05~\AA\ and 17.10 \AA\ line blend is optically thin is not completely correct, and our {\it{observed}} values of $(I/I_{0})_{15.01\AA}$ are biased slightly high, which further increases the significance of our upper limits on the turbulent velocities.
In Table~\ref{table:depths}, we also list the predicted suppression for the strongest emission lines, $(I/I_0)_{\mathrm{circ}}=\int I\mathrm{d}A\, /\int I_{0} \mathrm{d}A$, integrated within a circular region of radius 2~kpc, and within the RGS extraction region $(I/I_0)_{\mathrm{RGS}}$. The suppression for the RGS is determined in a simplified way, integrating within our 0.5\arcmin\ wide, and for line emission effectively about 3\arcmin\ long, RGS extraction region. These predicted ratios in Table~\ref{table:depths} were determined assuming a flat abundance distribution. By adopting an optically thin plasma in the analysis, the best fitting O abundance within the 2~kpc radius can be underestimated by $\sim$10--25\%, and the Fe abundance by $\sim$10--20\%. 

\section{Discussion and conclusions}

High-resolution spectra obtained with \xmm\ RGS reveal that the \ion{Fe}{xvii} line at 15.01~\AA\  in the cores of the elliptical galaxies NGC~4636, NGC~1404, NGC~5813, and NGC~4472 is suppressed by resonant scattering. The effects of resonant scattering can be investigated in detail for NGC~4636. 
The comparison of the measured suppression of the 15.01~\AA\ \ion{Fe}{xvii} line in the spectrum of NGC~4636 with the simulated effects of resonant scattering reveals that the characteristic velocity of isotropic turbulence in the core of the galaxy is smaller than 0.25 and 0.5 of the sound speed at the $\sim$90\% and 95\% confidence levels, respectively.  
The energy density in turbulence is:
\begin{equation}
\epsilon_{\mathrm{turb}}=\frac{3}{2}\, \rho\, V^2_{\mathrm{1D, turb}} = \frac{3}{4}\, \rho\, c_{\mathrm{s}}^2\, M^2
\end{equation}
where $\rho$ is the density of the plasma. The thermal energy of the plasma is given as:
\begin{equation}
\epsilon_{\mathrm{therm}}=\frac{3}{2}\frac{\rho}{\mu m_{\mathrm{p}}}kT.
\end{equation}
By combining these equations we obtain the ratio of turbulent to thermal energy:
\begin{equation}
\frac{\epsilon_{\mathrm{turb}}}{\epsilon_{\mathrm{therm}}}=\frac{\gamma}{2}\, M^2.
\end{equation}
For the 90\% and 95\% confidence level upper limits of $M\lesssim 0.25$ and 0.5 on the characteristic velocity of isotropic turbulence in NGC~4636, the upper limit on the fraction of energy in turbulent motions is 5\% and 20\%, respectively. This is in agreement with the upper limits on the non-thermal pressure component in elliptical galaxies obtained on larger spatial scales by \citet{churazov2008} and with the properties of AGN induced turbulence in three-dimensional simulations, that use adaptive mesh hydrodynamic code with a subgrid model of turbulence and mixing \citep{scannapieco2008}.

A small value for the average turbulent velocity dispersion was also inferred in the initial analysis of the RGS spectra for NGC~4636 by \citet{xu2002}. They estimated that the turbulent velocities in the galaxy do not strongly exceed 1/10 of the sound speed. \citet{caon2000} studied the kinematics of line emission nebulae in NGC~4636 and found an irregular radial velocity curve which indicates turbulent motions in the core. The measured radial velocities of the emission line nebulae are mostly  within about $\sim$100~km~s$^{-1}$.  The kinematics of emission line nebulae in the sample of elliptical galaxies studied by \citet{caon2000} indicates chaotic gas motions with velocities of about $M=0.2$--0.4. In cooling cores, optical emission line nebulae often spatially coincide with soft X-ray emission \citep[e.g.][]{sparks2004,fabian2006}, which indicates a coupling, suggesting that the two gas phases should share the same turbulent velocities. 

\begin{figure}
\includegraphics[width=0.350\textwidth,clip=t,angle=270.]{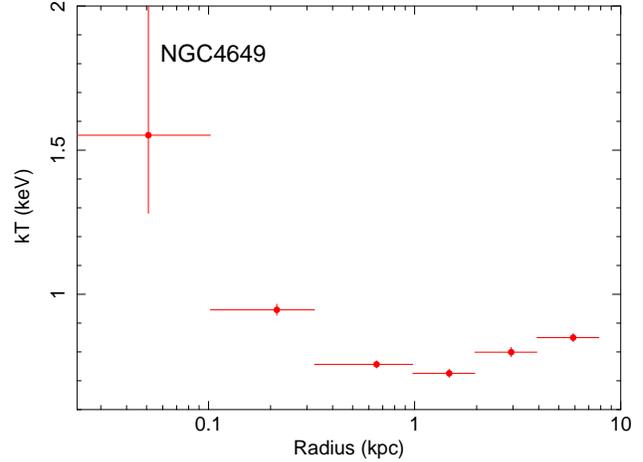}
\caption{The deprojected temperature profile of NGC~4649 determined using \chandra\ data. It shows an unusual central peak. } 
\label{ngc4649kTprof}
\end{figure}

The dense cores of NGC~4636 and NGC~5813 show strong evidence of recent violent interaction between the hot plasma and the AGN. This interaction is expected to induce turbulence through the mechanical activity of the outflows \citep{churazov2001}. Our results indicate that the typical velocities of the AGN induced turbulent motions in the hot plasma on sub-kpc spatial scales are less than 100--200~km~s$^{-1}$. Gas motions on larger spatial scales likely have larger velocities. Observations of the Perseus cluster suggest both turbulent and laminar gas motions with velocities as high as 700~km~s$^{-1}$ \citep{fabian2003b,fabian2003}. The observed lack of resonant scattering in the 6.7~keV Fe-K line within the inner $\sim$100~kpc \citep{churazov2004}  of Perseus confirms the presence of such strong gas motions. AGN induced gas motions with large range of velocities along our line of sight, perhaps even on spatial scales smaller than $\approx$1~kpc, could explain the observed lack of resonant scattering in the 15.01~\AA\ \ion{Fe}{xvii} line in M~87 \citep[the analysis of a 0.5\arcmin\ wide extraction region centred on the core of M~87 gives $(I_{\lambda17.05}+I_{\lambda17.10})/I_{\lambda15.01}=1.3\pm0.3$,][]{werner2006b}. We note that the central AGN in Perseus and in M~87 are more active in X-rays than the AGN in the target galaxies of this study. 

The dense cores of galaxies showing smaller disturbance by AGN might be expected to exhibit less turbulence than the heavily disturbed systems. Therefore, we might expect the 15.01~\AA\ line to be more suppressed in the most relaxed galaxies in our sample. While there is a weak indication of the suppression increasing from the least to the most relaxed systems: from NGC~5813 and NGC~4636, to NGC~4472, the large errorbars on the line ratios and the small number of galaxies in the sample do not allow us to draw firm conclusions. 

The best fit line ratio in the apparently relaxed galaxy NGC~4649 shows no indication for resonant scattering in its core. The lack of observable optical depth effects in this galaxy can be explained by the unusual temperature profile of its gaseous halo, which exhibits a sharp temperature peak in its centre (see Fig.~\ref{ngc4649kTprof}, Allen et al. in prep.). Within the central 0.1~kpc of the galaxy, the temperature increases by about a factor of 2 to $\sim$1.5 keV, at which no \ion{Fe}{xvii} is present in the gas, implying that all the observed \ion{Fe}{xvii} emission is from lower density regions surrounding the hot core. The high temperature might be due to shock heating by the central AGN \citep[e.g.][]{shurkin2008}. None of the other galaxies in our sample shows such an unusual temperature profile \citep[][Allen et al. in prep.]{allen2006}.

Unfortunately, the high temperatures of cluster cores do not allow us to use \ion{Fe}{xvii} lines as a diagnostic tool to measure the effects of resonant scattering and thus place firm constraints on turbulent velocities in the intra-cluster medium. No other lines have a comparable sensitivity to the turbulent velocities. However, the dense cores of elliptical galaxies, are in terms of the dominant physical processes like cooling and AGN feedback, sufficiently similar to cooling cores of clusters of galaxies to argue that the turbulent velocities on small spatial scales, and thus the turbulent pressure support in these systems, are also likely to be low. This is encouraging for cosmological studies based on X-ray observations of galaxy clusters \citep[e.g.][]{allen2008,mantz2008,henry2008,vikhlinin2008}.  Simulations predict that turbulent gas motions in the intra-cluster medium can provide 5\%--25\% of the total pressure support through the virial region of clusters, potentially biasing low hydrostatic mass estimates, even in relaxed systems \citep[e.g.][]{nagai2007}. However, these predictions depend sensitively on the gas physics assumed. Our results argue that such turbulence may be suppressed in nature, at least within the inner regions of such systems. For the merging unrelaxed Coma cluster, a $\sim$10\% lower limit on turbulent pressure support was obtained by \citet{schuecker2004} from a Fourier analysis of pressure maps. In the range 40--90~kpc, the turbulent spectrum was found to be well described by a projected Kolmogorov/Oboukhov-type model. 

Future, deeper (of the order of $\sim$500~ks) observations of large nearby elliptical galaxies with the \xmm\ RGS should provide even tighter constraints on turbulent velocities from resonant scattering. 
X-ray calorimeters on future missions such as {\it{Astro-H}} and {\it{International X-ray Observatory (IXO)}} will enable direct measurements of the velocity-broadening of emission lines in the hot gas in galaxies, groups, and clusters of galaxies \citep{sunyaev2003, inogamov2003, rebusco2008}, thereby allowing an independent check on our results as well as a critical extension to higher mass systems. However, we note that even the excellent (2--5~eV) spectral resolution of these instruments will not allow measurements of turbulent motions with velocities significantly smaller than 100--200~km~s$^{-1}$. Even though, the broad point spread function of the {\it{Astro-H}} mirrors will prohibit investigations of the effects of resonant scattering on spatial scales smaller than presented in this paper, it will for the first time allow us to directly measure the gas motions and spatially map them out to larger radii. The next planned X-ray observatory able to produce a real breakthrough by mapping the gas motions in groups and clusters of galaxies at high spatial resolution, will be {\it{IXO}}. 

An important consequence of resonant scattering is the bias it can introduce in measurements of radial profiles of abundances of chemical elements. As we show in Table~\ref{table:depths}, the integrated emission of several strong resonance lines is significantly suppressed within the inner 2~kpc of NGC~4636. By ignoring the effects of resonant scattering, the Fe and O abundance will be underestimated by 10--20\% within this region. This conclusion is in line with the results of \citet{sanders2006}. However, while resonant scattering makes abundances appear to be 10--20\% lower than the real value, it cannot explain the much stronger abundance dips seen in the cores of several clusters of galaxies \citep[e.g.][]{sanders2002}.

\section*{Acknowledgements}
Support for this work was provided by the National Aeronautics and Space Administration through Chandra Postdoctoral Fellowship Award Number PF8-90056
issued by the Chandra X-ray Observatory Center, which is operated by the Smithsonian Astrophysical Observatory for and on behalf of the National Aeronautics and 
Space Administration under contract NAS8-03060. The work of EC is supported by the DFG grant CH389/3-2. SWA acknowledges support from \chandra\ grant AR7-8007X. 
This work is based on observations obtained with \xmm, an ESA science mission with instruments
and contributions directly funded by ESA member states and the USA (NASA). This work was supported in part by the U.S. Department of Energy under contract number DE-AC02-76SF00515.

\bibliographystyle{aa}
\bibliography{clusters}

\begin{thebibliography}{42}
\expandafter\ifx\csname natexlab\endcsname\relax\def\natexlab#1{#1}\fi

\bibitem[{{Allen} {et~al.}(2006){Allen}, {Dunn}, {Fabian}, {Taylor}, \&
  {Reynolds}}]{allen2006}
{Allen}, S.~W., {Dunn}, R.~J.~H., {Fabian}, A.~C., {Taylor}, G.~B., \&
  {Reynolds}, C.~S. 2006, \mnras, 372, 21

\bibitem[{{Allen} {et~al.}(2008){Allen}, {Rapetti}, {Schmidt}, {Ebeling},
  {Morris}, \& {Fabian}}]{allen2008}
{Allen}, S.~W., {Rapetti}, D.~A., {Schmidt}, R.~W., {et~al.} 2008, \mnras, 383,
  879

\bibitem[{{Arnaud}(1996)}]{arnaud1996}
{Arnaud}, K.~A. 1996, in Astronomical Society of the Pacific Conference Series,
  Vol. 101, Astronomical Data Analysis Software and Systems V, ed. G.~H.
  {Jacoby} \& J.~{Barnes}, 17

\bibitem[{{Biller} {et~al.}(2004){Biller}, {Jones}, {Forman}, {Kraft}, \&
  {Ensslin}}]{biller2004}
{Biller}, B.~A., {Jones}, C., {Forman}, W.~R., {Kraft}, R., \& {Ensslin}, T.
  2004, \apj, 613, 238

\bibitem[{{Caon} {et~al.}(2000){Caon}, {Macchetto}, \& {Pastoriza}}]{caon2000}
{Caon}, N., {Macchetto}, D., \& {Pastoriza}, M. 2000, \apjs, 127, 39

\bibitem[{{Churazov} {et~al.}(2001){Churazov}, {Br{\"u}ggen}, {Kaiser},
  {B{\"o}hringer}, \& {Forman}}]{churazov2001}
{Churazov}, E., {Br{\"u}ggen}, M., {Kaiser}, C.~R., {B{\"o}hringer}, H., \&
  {Forman}, W. 2001, \apj, 554, 261

\bibitem[{{Churazov} {et~al.}(2004){Churazov}, {Forman}, {Jones}, {Sunyaev}, \&
  {B{\"o}hringer}}]{churazov2004}
{Churazov}, E., {Forman}, W., {Jones}, C., {Sunyaev}, R., \& {B{\"o}hringer},
  H. 2004, \mnras, 347, 29

\bibitem[{{Churazov} {et~al.}(2008){Churazov}, {Forman}, {Vikhlinin},
  {Tremaine}, {Gerhard}, \& {Jones}}]{churazov2008}
{Churazov}, E., {Forman}, W., {Vikhlinin}, A., {et~al.} 2008, \mnras, 388, 1062

\bibitem[{{den Herder} {et~al.}(2001){den Herder}, {Brinkman}, {Kahn},
  {Branduardi-Raymont}, {Thomsen}, {Aarts}, {Audard}, {Bixler}, {den Boggende},
  {Cottam}, {Decker}, {Dubbeldam}, {Erd}, {Goulooze}, {G{\" u}del},
  {Guttridge}, {Hailey}, {Janabi}, {Kaastra}, {de Korte}, {van Leeuwen},
  {Mauche}, {McCalden}, {Mewe}, {Naber}, {Paerels}, {Peterson}, {Rasmussen},
  {Rees}, {Sakelliou}, {Sako}, {Spodek}, {Stern}, {Tamura}, {Tandy}, {de
  Vries}, {Welch}, \& {Zehnder}}]{herder2001}
{den Herder}, J.~W., {Brinkman}, A.~C., {Kahn}, S.~M., {et~al.} 2001, \aap,
  365, L7

\bibitem[{{Doron} \& {Behar}(2002)}]{doron2002}
{Doron}, R. \& {Behar}, E. 2002, \apj, 574, 518

\bibitem[{{Fabian} {et~al.}(2003{\natexlab{a}}){Fabian}, {Sanders}, {Allen},
  {Crawford}, {Iwasawa}, {Johnstone}, {Schmidt}, \& {Taylor}}]{fabian2003}
{Fabian}, A.~C., {Sanders}, J.~S., {Allen}, S.~W., {et~al.} 2003{\natexlab{a}},
  \mnras, 344, L43

\bibitem[{{Fabian} {et~al.}(2003{\natexlab{b}}){Fabian}, {Sanders}, {Crawford},
  {Conselice}, {Gallagher}, \& {Wyse}}]{fabian2003b}
{Fabian}, A.~C., {Sanders}, J.~S., {Crawford}, C.~S., {et~al.}
  2003{\natexlab{b}}, \mnras, 344, L48

\bibitem[{{Fabian} {et~al.}(2006){Fabian}, {Sanders}, {Taylor}, {Allen},
  {Crawford}, {Johnstone}, \& {Iwasawa}}]{fabian2006}
{Fabian}, A.~C., {Sanders}, J.~S., {Taylor}, G.~B., {et~al.} 2006, \mnras, 366,
  417

\bibitem[{{Gastaldello} \& {Molendi}(2004)}]{gastaldello2004}
{Gastaldello}, F. \& {Molendi}, S. 2004, \apj, 600, 670

\bibitem[{{Gilfanov} {et~al.}(1987){Gilfanov}, {Sunyaev}, \&
  {Churazov}}]{gilfanov1987}
{Gilfanov}, M.~R., {Sunyaev}, R.~A., \& {Churazov}, E.~M. 1987, Soviet
  Astronomy Letters, 13, 3

\bibitem[{{Gonzal\'ez-Riestra}(2004)}]{riestra2004}
{Gonzal\'ez-Riestra}, R. 2004,
  http://xmm.vilspa.esa.es/docs/documents/CAL-TN-0058-1-0.ps.gz

\bibitem[{{Henry} {et~al.}(2008){Henry}, {Evrard}, {Hoekstra}, {Babul}, \&
  {Mahdavi}}]{henry2008}
{Henry}, J.~P., {Evrard}, A.~E., {Hoekstra}, H., {Babul}, A., \& {Mahdavi}, A.
  2008, ArXiv e-prints

\bibitem[{{Inogamov} \& {Sunyaev}(2003)}]{inogamov2003}
{Inogamov}, N.~A. \& {Sunyaev}, R.~A. 2003, Astronomy Letters, 29, 791

\bibitem[{{Jones} {et~al.}(2002){Jones}, {Forman}, {Vikhlinin}, {Markevitch},
  {David}, {Warmflash}, {Murray}, \& {Nulsen}}]{jones2002}
{Jones}, C., {Forman}, W., {Vikhlinin}, A., {et~al.} 2002, \apjl, 567, L115

\bibitem[{{Kaastra} {et~al.}(1996){Kaastra}, {Mewe}, \&
  {Nieuwenhuijzen}}]{kaastra1996}
{Kaastra}, J.~S., {Mewe}, R., \& {Nieuwenhuijzen}, H. 1996, in UV and X-ray
  Spectroscopy of Astrophysical and Laboratory Plasmas p.411, K. Yamashita and
  T. Watanabe. Tokyo : Universal Academy Press

\bibitem[{{Kalberla} {et~al.}(2005){Kalberla}, {Burton}, {Hartmann}, {Arnal},
  {Bajaja}, {Morras}, \& {P{\"o}ppel}}]{kalberla2005}
{Kalberla}, P.~M.~W., {Burton}, W.~B., {Hartmann}, D., {et~al.} 2005, \aap,
  440, 775

\bibitem[{{Lodders}(2003)}]{lodders2003}
{Lodders}, K. 2003, \apj, 591, 1220

\bibitem[{{Mantz} {et~al.}(2008){Mantz}, {Allen}, {Ebeling}, \&
  {Rapetti}}]{mantz2008}
{Mantz}, A., {Allen}, S.~W., {Ebeling}, H., \& {Rapetti}, D. 2008, \mnras, 387,
  1179

\bibitem[{{Markevitch} {et~al.}(2000){Markevitch}, {Ponman}, {Nulsen}, {Bautz},
  {Burke}, {David}, {Davis}, {Donnelly}, {Forman}, {Jones}, {Kaastra},
  {Kellogg}, {Kim}, {Kolodziejczak}, {Mazzotta}, {Pagliaro}, {Patel}, {Van
  Speybroeck}, {Vikhlinin}, {Vrtilek}, {Wise}, \& {Zhao}}]{markevitch2000}
{Markevitch}, M., {Ponman}, T.~J., {Nulsen}, P.~E.~J., {et~al.} 2000, \apj,
  541, 542

\bibitem[{{Nagai} {et~al.}(2007){Nagai}, {Vikhlinin}, \&
  {Kravtsov}}]{nagai2007}
{Nagai}, D., {Vikhlinin}, A., \& {Kravtsov}, A.~V. 2007, \apj, 655, 98

\bibitem[{{O'Sullivan} {et~al.}(2003){O'Sullivan}, {Ponman}, \&
  {Collins}}]{osullivan2003}
{O'Sullivan}, E., {Ponman}, T.~J., \& {Collins}, R.~S. 2003, \mnras, 340, 1375

\bibitem[{{Rebusco} {et~al.}(2008){Rebusco}, {Churazov}, {Sunyaev},
  {B{\"o}hringer}, \& {Forman}}]{rebusco2008}
{Rebusco}, P., {Churazov}, E., {Sunyaev}, R., {B{\"o}hringer}, H., \& {Forman},
  W. 2008, \mnras, 384, 1511

\bibitem[{{Reiprich} \& {B{\" o}hringer}(2002)}]{reiprich2002}
{Reiprich}, T.~H. \& {B{\" o}hringer}, H. 2002, \apj, 567, 716

\bibitem[{{Sanders} \& {Fabian}(2002)}]{sanders2002}
{Sanders}, J.~S. \& {Fabian}, A.~C. 2002, \mnras, 331, 273

\bibitem[{{Sanders} \& {Fabian}(2006)}]{sanders2006}
{Sanders}, J.~S. \& {Fabian}, A.~C. 2006, \mnras, 371, 1483

\bibitem[{{Scannapieco} \& {Br{\"u}ggen}(2008)}]{scannapieco2008}
{Scannapieco}, E. \& {Br{\"u}ggen}, M. 2008, \apj, 686, 927

\bibitem[{{Schuecker} {et~al.}(2004){Schuecker}, {Finoguenov}, {Miniati},
  {B{\"o}hringer}, \& {Briel}}]{schuecker2004}
{Schuecker}, P., {Finoguenov}, A., {Miniati}, F., {B{\"o}hringer}, H., \&
  {Briel}, U.~G. 2004, \aap, 426, 387

\bibitem[{{Shurkin} {et~al.}(2008){Shurkin}, {Dunn}, {Gentile}, {Taylor}, \&
  {Allen}}]{shurkin2008}
{Shurkin}, K., {Dunn}, R.~J.~H., {Gentile}, G., {Taylor}, G.~B., \& {Allen},
  S.~W. 2008, \mnras, 383, 923

\bibitem[{{Smith} {et~al.}(2001){Smith}, {Brickhouse}, {Liedahl}, \&
  {Raymond}}]{smith2001}
{Smith}, R.~K., {Brickhouse}, N.~S., {Liedahl}, D.~A., \& {Raymond}, J.~C.
  2001, \apjl, 556, L91

\bibitem[{{Sparks} {et~al.}(2004){Sparks}, {Donahue}, {Jord{\'a}n},
  {Ferrarese}, \& {C{\^o}t{\'e}}}]{sparks2004}
{Sparks}, W.~B., {Donahue}, M., {Jord{\'a}n}, A., {Ferrarese}, L., \&
  {C{\^o}t{\'e}}, P. 2004, \apj, 607, 294

\bibitem[{{Sunyaev} {et~al.}(2003){Sunyaev}, {Norman}, \&
  {Bryan}}]{sunyaev2003}
{Sunyaev}, R.~A., {Norman}, M.~L., \& {Bryan}, G.~L. 2003, Astronomy Letters,
  29, 783

\bibitem[{{Tamura} {et~al.}(2001){Tamura}, {Bleeker}, {Kaastra}, {Ferrigno}, \&
  {Molendi}}]{tamura2001}
{Tamura}, T., {Bleeker}, J.~A.~M., {Kaastra}, J.~S., {Ferrigno}, C., \&
  {Molendi}, S. 2001, \aap, 379, 107

\bibitem[{{Tamura} {et~al.}(2003){Tamura}, {Kaastra}, {Makishima}, \&
  {Takahashi}}]{tamura2003}
{Tamura}, T., {Kaastra}, J.~S., {Makishima}, K., \& {Takahashi}, I. 2003, \aap,
  399, 497

\bibitem[{{Vikhlinin} {et~al.}(2008){Vikhlinin}, {Kravtsov}, {Burenin},
  {Ebeling}, {Forman}, {Hornstrup}, {Jones}, {Murray}, {Nagai}, {Quintana}, \&
  {Voevodkin}}]{vikhlinin2008}
{Vikhlinin}, A., {Kravtsov}, A.~V., {Burenin}, R.~A., {et~al.} 2008, ArXiv
  e-prints

\bibitem[{{Vikhlinin} {et~al.}(2005){Vikhlinin}, {Markevitch}, {Murray},
  {Jones}, {Forman}, \& {Van Speybroeck}}]{vikhlinin2005}
{Vikhlinin}, A., {Markevitch}, M., {Murray}, S.~S., {et~al.} 2005, \apj, 628,
  655

\bibitem[{{Werner} {et~al.}(2006){Werner}, {B{\"o}hringer}, {Kaastra}, {de
  Plaa}, {Simionescu}, \& {Vink}}]{werner2006b}
{Werner}, N., {B{\"o}hringer}, H., {Kaastra}, J.~S., {et~al.} 2006, \aap, 459,
  353

\bibitem[{{Xu} {et~al.}(2002){Xu}, {Kahn}, {Peterson}, {Behar}, {Paerels},
  {Mushotzky}, {Jernigan}, {Brinkman}, \& {Makishima}}]{xu2002}
{Xu}, H., {Kahn}, S.~M., {Peterson}, J.~R., {et~al.} 2002, \apj, 579, 600

\end{thebibliography}

\end{document}